\documentclass{PoS}
\usepackage{bm}
\usepackage{amsthm}
\usepackage{wrapfig}
\title{A physicist-friendly reformulation of the Atiyah-Patodi-Singer index and its mathematical justification\footnote{The original title of the talk was ``Domain-wall fermion and Atiyah-Patodi-Singer index.''}}

\ShortTitle{A physicist-friendly APS index}

\author{
        \speaker{Hidenori Fukaya}$^a$\thanks{E-mail: hfukaya@het.phys.sci.osaka-u.ac.jp},
        Mikio Furuta$^{b}$\thanks{E-mail: furuta@ms.u-tokyo.ac.jp},
        Shinichiroh Matsuo$^{c}$\thanks{E-mail: shinichiroh@math.nagoya-u.ac.jp},
        Tetsuya Onogi$^{a}$\thanks{E-mail: onogi@het.phys.sci.osaka-u.ac.jp},
        Satoshi Yamaguchi$^{a}$\thanks{E-mail: yamaguch@het.phys.sci.osaka-u.ac.jp}, and
        Mayuko Yamashita$^{d}$\thanks{E-mail: mayuko@kurims.kyoto-u.ac.jp}
        \\
        \\
        \\
        \llap{$^a$}
        Department of Physics, Osaka University, Toyonaka, Japan
        \\
        \llap{$^b$}
        Graduate School of Mathematical Sciences, The University of Tokyo, Tokyo, Japan
        \\
        \llap{$^c$}
        Graduate School of Mathematics, Nagoya University, Nagoya, Japan
                \\
        \llap{$^d$}
        Research Institute for Mathematical Sciences, Kyoto University, Kyoto, Japan        
}

\abstract{The Atiyah-Patodi-Singer index theorem describes the bulk-edge correspondence of symmetry protected topological insulators.
  The mathematical setup for this theorem is, however, not directly related to the physical fermion system, as it imposes on the fermion
  fields a non-local and unnatural boundary condition known as the "APS boundary condition" by hand.
  In 2017, we showed that the same integer as the APS index can be obtained from the $\eta$ invariant of the domain-wall Dirac operator.
  Recently we gave a mathematical proof that the equivalence is not a coincidence but generally true.
  In this contribution to the proceedings of LATTICE 2019, we try to explain the whole story in a physicist-friendly way.
  }

\FullConference{37th International Symposium on Lattice Field Theory - Lattice2019\\
		16-22 June 2019\\
		Wuhan, China}

\begin{document}

\section{Introduction}

The Atiyah-Singer(AS) index theorem \cite{Atiyah:1968mp}
on a manifold without boundary is well-known
in physics, and it has played an important role
especially in high energy particle physics.
But its extension to the manifold with boundary,
known as Atiyah-Patodi-Singer(APS) index theorem \cite{Atiyah:1975jf},
was not discussed very much, as we were not that interested
in a space-time having boundaries.

Recently, the theorem is drawing attention from
the condensed matter physics.
This is because the APS index is a key to understand the bulk-edge
correspondence \cite{Hatsugai:1993ywa, Witten:2015aba} of topological insulators
from anomaly matching of the time-reversal(T) symmetry.

It is, however, difficult to understand why
the APS index ``must'' appear in the physics of
the topological insulators,
since the original set up of the APS index
employs a  very unnatural boundary condition
and is unlikely to be realized
in the real electron systems.

In \cite{Fukaya:2017tsq}, three physicist half of the authors found a
different fermionic quantity, which coincides with the APS index.
We used a domain-wall Dirac operator \cite{Jackiw:1975fn, Callan:1984sa, Kaplan:1992bt}
on a closed manifold without boundary,
sharing a ``half'' of it with the original set up of the APS.
We have perturbatively shown that its  $\eta$ invariant, defined as a regularized difference of the number of
positive and negative eigenmodes, is equal to the original APS index.
Since the domain-wall Dirac fermion shares properties
of the electron systems of the topological insulators, 
we proposed this $\eta$ invariant
as a ``physicist-friendly'' reformulation of the APS index.

In \cite{Fukaya:2019qlf}, three mathematician half of the authors joined
the collaboration, and we succeeded in a proof that the above
observation of \cite{Fukaya:2017tsq} is not a coincidence but generally true.
Namely, we proved that for any APS index on a manifold with boundary
(we denote $X_+$ and its boundary $Y$), there exists a domain-wall Dirac operator
on a closed manifold without boundary, where its half coincides with $X_+$,
and its $\eta$ invariant is equal to the original APS index.

The key of this work is to add a mass term to the Dirac operator.
The mass term breaks the chiral symmetry, which is apparently
an essential property to describe the index theorems.
Nevertheless, as we will show below, there is no problem
in giving a fermionic integer, which coincides with the original index.
This is true even on a lattice, and we proposed a
non-perturbative formulation of the APS index in lattice gauge theory 
in \cite{Fukaya:2019myi}. In fact, the $\eta$ invariant of the massive Dirac operator
gives a unified view of the index theorems including their lattice version.
See also Kawai's contribution \cite{Kawai} to these proceedings.

\section{Why APS index unphysical?}

First we review the original work of APS and
discuss how it is unnatural.
We consider a Dirac operator
$
D = \sum_\mu \gamma_\mu \left(\frac{\partial}{\partial x_\mu}+iA_\mu(x)\right),
$
which operates on fermion fields on
a four-dimensional Euclidean flat space in the $x_4>0$ region only\footnote{
In order to make the index well-defined, we have to compactify the space time.
But here we perform our computation as if we were in a semi-infinite space
for simplicity of the presentation.}.
Here, $A_\mu$ is the $SU(N)$ or $U(1)$ gauge field and
we take $A_4=0$ gauge, which simplifies our computation.

The APS boundary condition is given by the operator at $x_4=0$,
\begin{eqnarray}
  B &=& \gamma_4 \sum_{i=1}^3 \gamma_i \left.\left(\frac{\partial}{\partial x_i}+iA_i(x)\right)\right|_{x_4=0},
\end{eqnarray}
such that any positive eigencomponent of $B$ becomes zero at $x_4=0$,
which guarantees the Hermiticity of the Dirac operator.
As $[B,\gamma_5]=0$, the chiral symmetry is conserved.
In fact, $B$ can be block-diagonal: $B={\rm diag}(iD^{\rm 3D}, -iD^{\rm 3D})$,
where $iD^{\rm 3D}$ is the three-dimensional Dirac operator on the surface.
Since this condition requires information of the whole
eigenfunctions of $B$, it is non-local.

Let us compute the axial $U(1)$ anomaly
of a massless fermion with the APS boundary condition.
The path integral measure transforms as
$
d\psi d\bar{\psi}\to d\psi d\bar{\psi}\exp\left[2i \alpha {\rm Tr}\gamma_5\right],
$
under the chiral transformation.
The trace can be evaluated by the heat-kernel regulator as
\begin{eqnarray}
  \label{eq:index}
        {\rm Tr} \gamma_5 e^{D^2/M^2}\equiv \lim_{M\to \infty}\int d^4 x {\rm tr}\gamma_5 e^{D^2/M^2}.
\end{eqnarray}
If it were without boundary, the plane wave complete set would apply to obtain the AS index.
However, we need here a different complete set, which satisfies the
APS boundary condition.

At the leading order of the adiabatic expansion in $x_4$,
let us consider the eigenproblem
of $-D^2\psi(x)=\Lambda^2 \psi(x)$, where the solution is given by
$\psi = \phi_\pm(x_4)\otimes \phi_\lambda^{\rm 3D}(\bm{x})$, choosing
the three-dimensional part $\phi_\lambda^{\rm 3D}(\bm{x})$ 
to be the eigenfunction of $iD^{\rm 3D}$ with the eigenvalue $\lambda$.
The results are
\begin{eqnarray}
  \phi_\pm^p(x_4) &=& \frac{u_\pm}{\sqrt{2\pi}}\left(e^{i\omega x_4}-e^{-i\omega x_4}\right),\;\;\;
  \phi_\pm^n(x_4) = \frac{u_\pm\left((i\omega \mp \lambda)e^{i\omega x_4}-c.c.\right)}{\sqrt{2\pi(\omega^2+\lambda^2)}},
\end{eqnarray}
where $\phi_\pm^p(x_4)$ denotes the positive eigenmodes of $B$,
while $\phi_\pm^n(x_4)$ is the negative modes. We take $\gamma_5 u_\pm = \pm u_\pm$
and $\omega^2=\Lambda^2-\lambda^2$, which must be positive so that there is no edge mode allowed.

Using the above complete set, Eq.~(\ref{eq:index}) is evaluated as
\begin{eqnarray}
  \sum_\lambda \int_{x_4>0} dx_4 {\rm sgn}\lambda e^{-\lambda^2/M^2}
  \int \frac{d\omega}{2\pi} \left(-1 +\frac{2i|\lambda|}{\omega+i|\lambda|}\right)e^{-\omega^2/M^2+2i\omega x_4}
  = - \sum_\lambda \frac{{\rm sgn}\lambda }{2}
  {\rm erfc}(|\lambda|/M),
\end{eqnarray}
which gives the $\eta$ invariant of $iD^{\rm 3D}$ in the $M\to \infty$ limit.
From the next-to-leading order contribution of the adiabatic expansion, 
we obtain the curvature term in the $x_4>0$ region.
In total, we have
\begin{eqnarray}
  \label{eq:APSorg}
  {\rm Ind}_{\rm APS}(D)= {\rm Tr} \gamma_5 e^{D^2/M^2} = 
  \frac{1}{32\pi^2}\int_{x_4>0} d^4x\; \epsilon_{\mu\nu\rho\sigma}{\rm tr}F^{\mu\nu}F^{\rho\sigma}-\frac{\eta(iD^{\rm 3D})}{2}.
\end{eqnarray}

The above derivation has no problem in mathematics but
physically unnatural. 
First of all, the causality becomes questionable with the non-locality\footnote{
In a recent work \cite{Witten:2019bou},  the use of  the APS boundary condition is justified by rotating
the boundary to the temporal direction and regarding it as an intermediate state
in the partition function of massive fermion systems.
Here we give an alternative way, which allows us to take the boundary remaining in space-like direction.
}, as any change in the eigenfunction of $B$ is immediately reflected
to the whole boundary, which means that the information propagates faster than speed of light.
Second, as $B$ involves the momentum and gauge fields in spatial directions,
the APS boundary condition does not respect the rotational symmetry on the surface.
Third problem is the fact that there is no edge-localized mode allowed to exist under the APS condition.

This unnaturalness of the APS boundary condition motivated us to
explore a physicist-friendly reformulation of the APS index.
What we have to give up is clear when we consider the reflection
of the fermionic particle at the boundary.
As the translational invariance is lost in the $x_4$ direction,
the momentum in that direction is not conserved.
But as the energy has to be conserved 
the fourth momentum must flip at the boundary.
On the other hand, the angular momentum in the $x_4$ direction
should be conserved, which means that the helicity must flip by the reflection.
It is, therefore, essential to give up the chirality and we have to consider
the index theorem with massive fermions.

\section{Atiyah-Singer index in terms of the massive Dirac operator}

Here let us consider a simpler case, an Euclidean four-dimensional space
without boundary, and try to reformulate the AS index in terms of the
massive Dirac operator.
The Dirac fermion partition function is given by
\begin{equation}
  \label{eq:massivedet}
\det \frac{D+M}{D+M_2},
\end{equation}
where we have introduced the Pauli-Villars regulator
with a mass $M_2\gg M$.
When $M$ and $M_2$ have the same sign,
the large $M\to M_2$ limit leads to a trivial consequence that
the above determinant is unity,
which is the situation of normal insulators.

When the sign of the mass is flipped, the situation is different.
For the determinant
\begin{equation}
  \label{eq:nmassivedet}
\det \frac{D-M}{D+M_2},
\end{equation}
one can flip the sign of the mass by the chiral rotation with $\alpha=\pi$, but
the measure changes as
\begin{eqnarray}
  \label{eq:measure}
  \det \frac{D-M}{D+M_2}  
  \propto \exp\left(i\pi \frac{1}{32\pi^2}\int d^4x\; \epsilon_{\mu\nu\rho\sigma}{\rm tr}F^{\mu\nu}F^{\rho\sigma}\right) 
  = \exp(i\pi Q) = (-1)^Q,
\end{eqnarray}
which means a nontrivial $\theta=\pi$ vacuum. In fact, $Q$ is the AS index.

The same determinant can be evaluated in a different way as
\begin{eqnarray}
  &\det \frac{D-M}{D+M_2} = \det \frac{i\gamma_5(D-M)}{i\gamma_5(D+M_2)} = \frac{\prod i\lambda(-M)}{\prod i\lambda(M_2)}
  \propto \exp \left[i\pi \left( \frac{\sum{\rm sgn}\lambda(-M)}{2}- \frac{\sum{\rm sgn}\lambda(M_2)}{2}\right)  \right],
\end{eqnarray}
where $\lambda(m)$ is the eigenvalue of $\gamma_5(D+m)$.
Note that the exponent is nothing but an expression of the $\eta$ invariant (with the Pauli-Villars subtraction),
and therefore, we can conclude
\begin{equation}
  \label{eq:ASeta}
  - \frac{\eta^{PV reg.}(\gamma_5(D-M))}{2} = Q,
\end{equation}
which coincides with the AS index. For more mathematical proof, see our paper \cite{Fukaya:2019qlf}.


\section{APS index and domain-wall fermion}

Now let us consider the domain-wall fermion
determinant,
\begin{equation}
  \label{eq:detDW}
\det \frac{D-M\varepsilon(x_4)}{D+M_2},
\end{equation}
where $\varepsilon(x)=x/|x|$ is the sign function.
Unlike the original set up by APS, we include
the $x_4<0$ region in the system.
Because of the $\gamma_5$ Hermiticity:
$D^\dagger =\gamma_5 D\gamma_5$, the determinant
is still real, and therefore, the sign of
the determinant is controlled by an integer $Q$.

Note that the physical properties
of the domain-wall fermion are similar to those
of topological insulators.
The fermion is massive(gapped) in the bulk, while the
massless(gapless) mode appears at the wall between
two physically different regions.
We also note that any topological insulator in our
world is surrounded by normal insulators.
Therefore, it is more natural to model the physics of
topological insulators with this domain-wall fermion
than the one on a manifold with boundary.

The domain-wall fermion determinant in Eq.~(\ref{eq:detDW})
can be expressed as
\begin{eqnarray}
  = \det \frac{i\gamma_5(D-M\varepsilon(x_4))}{i\gamma_5(D+M)}
  \propto \exp \left[ -i\pi \frac{\eta^{PV reg.}(H_{DW})}{2}\right],
\end{eqnarray}
where $H_{DW}=\gamma_5(D-M\varepsilon(x_4)))$.
Therefore, we can conclude
\begin{equation}
  \label{eq:APSeta}
  - \frac{\eta^{PV reg.}(H_{DW})}{2} = Q.
\end{equation}
In fact, this integer coincides with the APS index.

In our paper \cite{Fukaya:2017tsq}, we evaluated the $\eta$ invariant,
using the integral expression for $1/\sqrt{H_{DW}^2}$ as
\begin{eqnarray}
\frac{1}{\sqrt{H_{DW}^2}} = \frac{1}{\sqrt{\pi}}\int_0^\infty dt t^{-1/2}e^{-t H_{DW}^2},
\end{eqnarray}
and expanding the exponential part $e^{-t H_{DW}^2}$ in the gauge coupling.
The essential point here is that even in the zero coupling limit,
$H_{DW}^2$ has a non-trivial structure as
\begin{eqnarray}
  H_{DW}^2 = -D^2 + M^2 - 2 M \gamma_4 \delta(x_4),
\end{eqnarray}
where the last delta-function comes from the domain-wall.
In fact, it has an edge-localized mode, whose eigenfunction
in the $x_4$ direction is
\begin{eqnarray}
\phi(x_4)= C \exp(-M |x_4|), \;\; \gamma_4\phi(x_4)=-\phi(x_4),
\end{eqnarray}
and from this edge mode, we have reproduced the second term of Eq.~(\ref{eq:APSorg}).

On the other hand, the extended modes have higher energy than $M$,
and therefore, their contribution to the
$\eta$ invariant reduces to the integral of a local quantity
that is exactly the first term of Eq.~(\ref{eq:APSorg}).
It is important to note that
we did not assume any boundary condition at $x_4=0$ but the delta function
potential automatically chooses a local and rotationally symmetric
boundary condition.

\section{Mathematical proof}

Recently, the mathematician-half of the authors joined
and we succeeded in proving the equivalence
of the APS index and the domain-wall $\eta$ invariant on a general even-dimensional manifold  \cite{Fukaya:2019qlf}. The main theorem we have proved is
\newtheorem{FFMOYY}{Theorem}

\begin{FFMOYY}
  Let $X$ be a $2n$-dimensional closed and oriented manifold
  and $S$ be a Hermitian vector bundle (whose section corresponds to the fermion field) on $X$.
  We assume that $S$ is $\mathbb{Z}_2$ graded and $\Gamma_S$ as its $\mathbb{Z}_2$ grading operator
  (or $\gamma_5$).
  Let $D$ be a first-order and elliptic partial differential operator (or Dirac operator multiplied by $\gamma_5$),
  which anti-commutes with $\Gamma_S$. Let $Y$ be a separating sub-manifold that
  decomposes $X$ into two compact manifolds $X_+$ and $X_-$ ($Y$ is the domain-wall).
  We consider a step function $\kappa$, which takes $\pm 1$ on $X_\pm$
  (or $\varepsilon(x_4)$ in the previous section).

  Then, there exists $m_0>0$ and for any $m>m_0$,
\begin{equation}
  \label{eq:maintheorem}
  {\rm Ind}_{\rm APS}(D|_{X_+}) = -\frac{\eta(D-m\kappa \Gamma_S)-\eta(D+m\Gamma_S)}{2}
\end{equation}
holds.
Here, the left-hand side is the APS index on $X_+$ with the APS boundary condition on $Y$.
The right-hand side denotes the $\eta$ invariant of the domain-wall fermion Dirac operator,
regularized by the Pauli-Villars fields, where we have chosen $M=M_2=m$.
\end{FFMOYY}

Below we give a rough sketch of the proof,
for which we need three known mathematical theorems:
1) The APS index is equal to the AS index on a manifold with an infinite cylinder
attached to the original boundary, where the gauge fields and metric are constant
on the cylinder. 2) Localization and product formula \cite{Witten:1982im, Furuta}:
adding a potential term,
we can ``localize'' the zero mode eigenfunction in a vicinity of a lower-dimensional
sub-manifold, and we can evaluate the index as a product of
the index in lower dimension and that in the normal direction.
3) The APS index on an odd-dimensional manifold is expressed by the boundary $\eta$ invariant only.

Here we introduce a manifold $\mathbb{R}\times X$ and $t$ as the coordinate in the extra dimension.
Then we consider the following operator  
\begin{eqnarray}
  \bar{D} = \left(
  \begin{array}{cc}
    0 &(D-m\rho\Gamma_S)+\partial_t\\
    (D-m\rho\Gamma_S)-\partial_t & 0
  \end{array}  
  \right),\;\;\; \rho = \left\{
  \begin{array}{cc}
  1 & \mbox{on $[0,+\infty]\times X_+$}\\
  -1 & \mbox{otherwise}
    \end{array}\right..
\end{eqnarray}
Note that $\rho=\kappa$ at $t=+1$, and $\rho=-1$ at $t=-1$.

\begin{wrapfigure}{r}{0.3\textwidth}
\begin{center}
  \includegraphics[width=0.4\textwidth, angle=90, bb= 0 0 737 460]{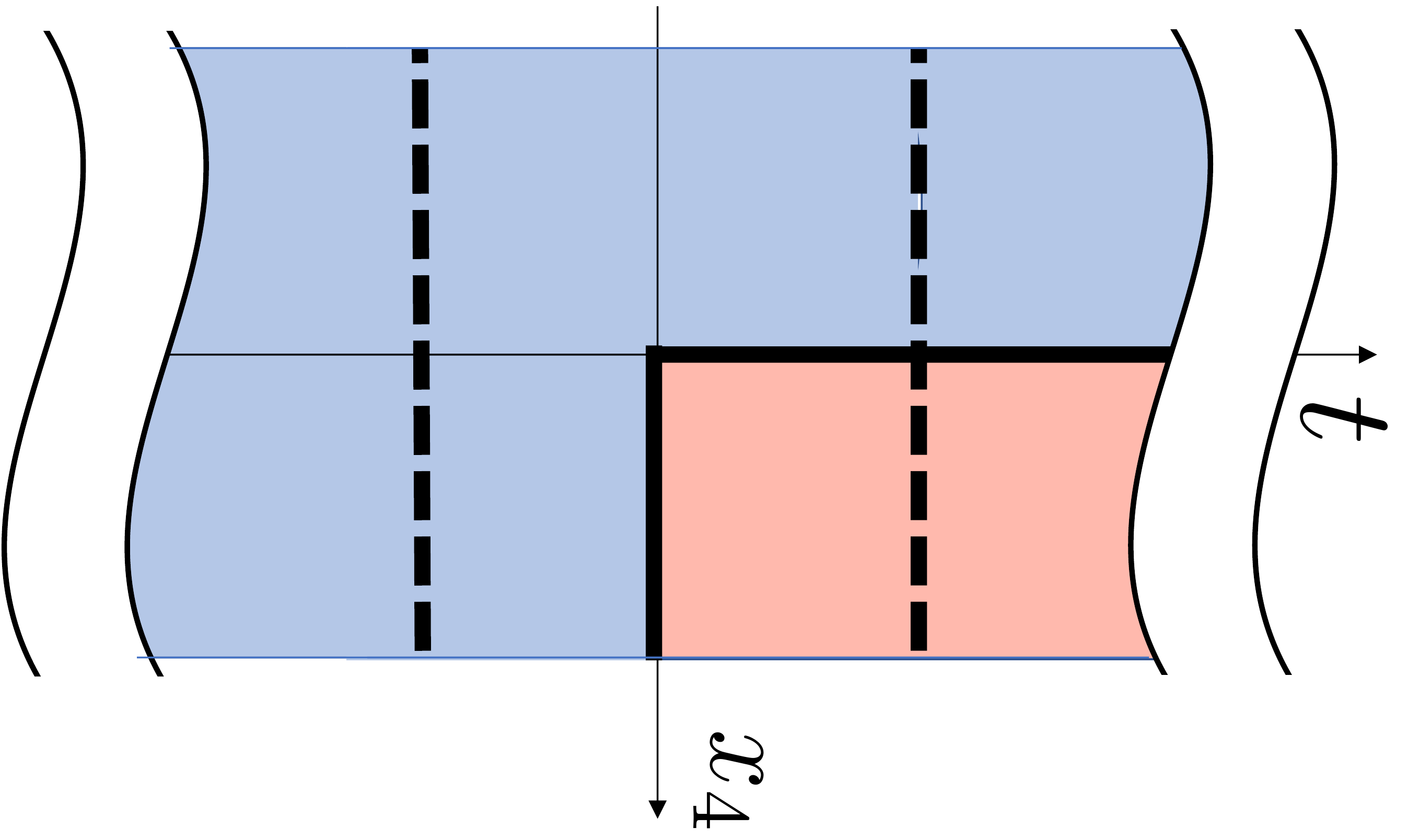}
\end{center}
\caption{
  Schematic picture of the manifold $\mathbb{R}\times X$.
  The horizontal axis represents the original four-dimensional $X$,
  and $X_+$ is shown in the $x_4>0$ region. The original figure
  was given in \cite{SK}.
}
\label{fig:5DDW}
\end{wrapfigure}
Let us evaluate the index of $\bar{D}$ in two different ways.
First, we use the theorem 2) taking the large $m$ limit, then
the zero eigenmodes of $\bar{D}$ are localized on a slice, which consists of
$X_+$ at $t=0$ and a cylinder $[0,\infty)\times Y$ (solid lines in Fig.~\ref{fig:5DDW}).
  Using the theorem 1) we can cut off the cylinder part and obtain
  ${\rm Ind}(\bar{D})={\rm Ind}_{\rm APS}(D|_{X_+})$.
  The second evaluation starts from the theorem 1) to cut down
  $\mathbb{R}\times X$ to $[-1,1]\times X$,
  putting the APS boundary condition
  on $t=\pm 1$(dashed lines in Fig.~\ref{fig:5DDW}).
  Then from the theorem 3), we obtain
  \begin{eqnarray*}
  {\rm Ind}(\bar{D})= {\rm Ind}_{\rm APS}(\bar{D}|_{[-1,1]\times X}) 
  = -\frac{\eta(D-m\kappa \Gamma_S)-\eta(D+m\Gamma_S)}{2},
  \end{eqnarray*}
  which is the right-hand side of Eq.~(\ref{eq:maintheorem}).
  Namely, Eq.~(\ref{eq:maintheorem}) always holds since its both sides are
  just two different expressions of the same index ${\rm Ind}(\bar{D})$.

\section{Summary and discussion}

We have shown that the $\eta$ invariant of the domain-wall fermion Dirac operator
gives a physicist-friendly reformulation of the APS index.
Using a Dirac operator in a higher dimensions,
we have given a mathematical proof that this reformulation is valid
on any even-dimensional curved manifold.
As the domain-wall fermion shares similar properties to those
of topological insulators, we believe that our work gives a
mathematical foundation to describe the bulk-edge correspondence of
topological matters.

The proof of the equivalence used ${\rm Ind}(\bar{D})$ on $\mathbb{R}\times X$
or equivalently,  ${\rm Ind}_{\rm APS}(\bar{D})$ on $[-1,1]\times X$.
An interesting extension is to express this index by an $\eta$ invariant again,
\begin{equation}
  {\rm Ind}_{\rm APS}(\bar{D}|_{[-1,1]\times X}) = -\frac{1}{2}\eta(\bar{D}-\bar{m}\bar{\kappa}\bar{\Gamma})^{reg.},
\end{equation}
where we have introduced a second mass term $\bar{m}\bar{\Gamma}={\rm diag}(\bar{m}\mathbb{I}_S,-\bar{m}\mathbb{I}_S)$,
with $\mathbb{I}$ an identity operator on $S$, and $\bar{\kappa}$ taking $\bar{\kappa}=1$ in $t \in [-1,1]$,
and $-1$, otherwise.
Then the original edge mode localized in $2n-1$ dimensional manifold $Y$ becomes
the edge-of-edge states of $\bar{D}-\bar{m}\bar{\kappa}\bar{\Gamma}$,
which is localized at the junction of the first and second domain-walls.
This recursive structure might be useful for the physics of higher order topological insulators.

The authors thank the organizers of the workshop Progress in the Mathematics
of Topological States of Matter, which triggered our collaboration.
This work was supported in part by JSPS KAKENHI (Grant numbers: JP15K05054, JP17H06461, JP17K14186, JP18H01216, JP18H04484, JP18K03620, and 19J22404).

\end{document}